\documentstyle[]{mn}

\begin{document}

%BEGIN USER-DEFINED SHORTFORMS
\def\spose#1{\hbox to 0pt{#1\hss}}
\def\simlt{\mathrel{\spose{\lower 3pt\hbox{$\mathchar"218$}}
     \raise 2.0pt\hbox{$\mathchar"13C$}}}
\def\simgt{\mathrel{\spose{\lower 3pt\hbox{$\mathchar"218$}}
     \raise 2.0pt\hbox{$\mathchar"13E$}}}
\def\eg{{\rm e.g., }}
\def\ie{{\rm i.e. }}
\def\etal{{\rm et~al. }}

\def\aj{{AJ}}			
\def\araa{{ARA\&A}}		
\def\apj{{ApJ}}			
\def\apjs{{ApJS}}		
\def\apss{{Ap\&SS}}		
\def\aap{{A\&A}}		
\def\aapr{{A\&A~Rev.}}		
\def\aaps{{A\&AS}}		
\def\azh{{AZh}}			
\def\jrasc{{JRASC}}		
\def\mnras{{MNRAS}}		
\def\pasa{{PASA}}		
\def\pasp{{PASP}}		
\def\pasj{{PASJ}}		
\def\sovast{{Soviet~Ast.}}	
\def\ssr{{Space~Sci.~Rev.}}	
\def\zap{{ZAp}}			
\def\nat{{Nature}}		
\def\aplett{{Astrophys.~Lett.}}
\def\fcp{{Fund.~Cosmic~Phys.}}
\def\memsai{{Mem.~Soc.~Astron.~Italiana}}
\def\nphysa{{Nucl.~Phys.~A}}
\def\physrep{{Phys.~Rep.}}
%END USER-DEFINED SHORTFORMS

\title[Supernovae and Intracluster Medium Enrichment]{Supernovae Types 
Ia/II and Intracluster Medium Enrichment}
\author[B.K. Gibson, M. Loewenstein \& R.F. Mushotzky]{B.K. Gibson$^1$, 
M. Loewenstein$^2$ \& R.F. Mushotzky$^2$\\
$^1$Mount Stromlo \& Siding Spring Observatories, Australian National
University, Weston Creek P.O., Weston, ACT, Australia  2611\\
$^2$Laboratory for High Energy Astrophysics, NASA/GSFC, Code 662, Greenbelt,
Maryland, USA  20771}
\maketitle
\begin{abstract}
We re-examine the respective roles played by supernovae (SNe) Types Ia and
II in enriching the intracluster medium (ICM) of galaxy clusters, in light of
the recent downward shift of the ASCA abundance ratios of $\alpha$-elements
to iron favoured by Ishimaru \& Arimoto (1997, PASJ, 49, 1).  
Because of this shift, Ishimaru \&
Arimoto conclude that $\simgt 50$\% of the ICM iron must have originated
from within Type Ia SNe progenitors.  A point not appreciated in their study,
nor in most previous analyses, is the crucial dependence of 
such a conclusion upon the adopted massive star physics.  Employing several
alternative Type II SN yield compilations, we demonstrate how uncertainties in
the treatment of convection and mass-loss can radically alter our perception of
the relative importance of Type Ia and II SNe as ICM polluters.  If mass-loss
of the form favoured by Maeder (1992, A\&A, 264, 105) or convection of the form
favoured by Arnett (1996, Supernovae and Nucleosynthesis) is assumed, the
effect
upon the oxygen yields would lead us to conclude that Type Ia SNe play no part
in polluting the ICM, in contradiction with Ishimaru \& Arimoto.  
Apparent dichotomies still exist (\eg the mean ICM
neon-to-iron ratio implies a $\sim 100$\% Type II Fe origin, while the mean
sulphur ratio
indicates a $\sim 100$\% Type Ia origin) that cannot be reconciled with the
currently available yield tables.
\end{abstract}


\section{Introduction}
\label{introduction}

Evidence for supersolar abundance ratios of $\alpha$-elements to iron (\ie
[$\alpha$/Fe]$\simgt +0.0$)
in the intracluster medium (ICM) of the Virgo, Perseus, and Abell 576 galaxy
clusters has been available in the literature since the mid-1980s (Canizares
\etal 1982, Canizares, Markert \& Donahue 1988, and Rothenflug \etal 1984),
although these measurements were heavily weighted towards thermally-complex
cooling flow regions.
Much debate in the subsequent years has
centred on the robustness of these results, in light of the difficulty in
obtaining accurate abundances for the $\alpha$-elements.  Taking advantage of
the unique characteristics of the \it
Advanced Satellite for Cosmology and Astrophysics \rm (ASCA),
Mushotzky \etal (1996) determined the ICM abundance ratios for four clusters
and found, using the solar photospheric abundance scale, the following
unweighted mean (SIS) values: [O/Fe]$\approx$+0.18, [Si/Fe]$\approx$+0.31,
[Ne/Fe]$\approx$+0.29, [S/Fe]$\approx$-0.11, and [Mg/Fe]$\approx$+0.07.
Scaling to the meteoritic abundances, Ishimaru \& Arimoto (1997) revised
these values to: [O/Fe]$\approx$+0.01, [Si/Fe]$\approx$+0.14,
[Ne/Fe]$\approx$+0.12, [S/Fe]$\approx$-0.32, and [Mg/Fe]$\approx$-0.10.
Similar results are becoming available for $\sim 10$ additional clusters.
Both ``scalings'' were based upon Table 1 of Anders \& Grevesse (1989).

The abundance ratio
pattern in the ICM provides a unique tool with which to probe the
origin of these heavy elements.  While the favoured mechanism for enriching the
ICM is supernovae (SNe)-driven winds from elliptical galaxies (\eg Gibson 1997,
and references therein), there is still no consensus as to whether these winds
are dominated by the ejecta of Type II SNe (either via initial mass functions
heavily-weighted towards their progenitors or via an early wind relatively
unpolluted by the longer-lived progenitors to Type Ia SNe) or Type Ia SNe.  
Since \it a
priori \rm galactic wind models can be constructed to satisfy either
scenario (\eg Matteucci \& Vettolani 1988; Gibson \& Matteucci 1997), it was
recognised that accurate cluster ICM abundance determinations might allow
discrimination between these two competing scenarios (simply by comparing 
with the abundance
ratio pattern for any Type II or Type Ia SNe predominance).

Recently Ishimaru \& Arimoto (1997), utilizing the meteoritic abundance
scaling and Type II SN yields from Tsujimoto \etal (1995),
concluded that $\simgt 50$\% of the ICM Fe
originated from Type Ia SNe.
Loewenstein \& Mushotzky (1996), using self-consistent arguments independent
of the assumed solar abundance scaling
(despite implications to the contrary
in Ishimaru \& Arimoto 1997), showed that models where all of the enrichment
was due to Type II SN were consistent with the ASCA data -- although they
took care to note that a significant Type Ia  SNe contribution to the
Fe enrichment could not be ruled out. The important difference in their
models was the adoption of Type II SN yields from Woosley \& Weaver (1995).
It is our
goal herein 
to more fully and systematically
examine how these Type Ia versus Type II ICM iron contribution arguments are
\it crucially \rm dependent upon the adopted theoretical
Type II SNe yields.
Such an
appreciation was not apparent in Ishimaru \& Arimoto (1997),
who erroneously attributed their relatively high inferred
Type Ia SNe contribution to
the ICM Fe enrichment to
their rescaling of
the original Mushotzky \etal (1996) abundance ratio
determinations; in fact their
conclusions were tied inexorably to the Tsujimoto \etal (1995) yields.

In Section \ref{analysis}, we present the basic framework necessary to
replicate the Ishimaru \& Arimoto (1997) analysis.  Instead of
restricting ourselves to a single Type II SNe yield compilation, though,
we consider
several competitors.  We list their inherent differences (\eg
differing treatments of convection, mass-loss, reaction rates) and concentrate
on their effects upon the predicted ICM elemental abundance ratios.
In what follows, we are solely interested in abundance \it ratio \rm questions,
and thus for simplicity's sake we shall simply adopt the Salpeter
(1955) IMF used by Ishimaru \& Arimoto.  Arguments pertaining to \it absolute
\rm abundance masses favour an IMF biased toward massive stars (\eg
Loewenstein \& Mushotzky 1996; Gibson \& Matteucci 1997), but these are not
particularly relevant to the analysis that follows.  

Will we have a definitive answer to the question of
Type Ia versus Type II ICM iron origin
at the conclusion of this paper?  To anticipate our conclusions, no.  What we
hope to leave with the reader is a better appreciation of the uncertainties
involved, and in particular, how a definitive answer {\it cannot} 
be reached until
further convergence of Type II SNe models is achieved.
Our results are summarised in Section \ref{summary}.

\section{Analysis}
\label{analysis}

In order to quantify the roles played by SNe Types Ia and II in contributing to
the ICM heavy element abundances, we make use of the formalism presented in
Ishimaru \& Arimoto (1997).  Let us restrict ourselves, for the time being, to
the question of the ICM iron origins.  We write the fractional contribution of
SNe Type Ia to the ICM iron as
\begin{equation}
{{\rm M_{\rm Fe,SNIa}}\over{\rm M_{\rm Fe,total}}} = {{\zeta y_{\rm Fe,SNIa}}
\over{\zeta y_{\rm Fe,SNIa}+(1-\zeta)<y_{\rm Fe,SNII}>}},
\label{eq:ICMfe}
\end{equation}
\noindent
where $\zeta$ is the frequency of Type Ia 
relative to Type II SNe.  The iron yield 
averaged over the Type II SN progenitor IMF, $\phi\propto m^{-x}$, is written
\begin{equation}
<y_{\rm Fe,SNII}> = {{\int_{m_\ell}^{m_u}y_{\rm Fe,SNII}(m)\phi(m)m^{-1}{\rm
d}m}\over{\int_{m_\ell}^{m_u}\phi(m)m^{-1}{\rm d}m}}.
\label{eq:SNIIfe}
\end{equation}
\noindent
The bounds for Type II SNe progenitors are taken to be $m_\ell=10$ M$_\odot$
and $m_u=50$ M$_\odot$, respectively, and an IMF slope $x=1.35$ (Salpeter 1955)
is adopted throughout. We have purposefully avoided considering 
$x$ as an additional free parameter in the analysis.  
While this slope is of prime importance for
arguments concerning the absolute mass of elements in the ICM (\eg Loewenstein
\& Mushotzky 1996), it is less so for abundance \it ratios\rm. 

The primary ingredients in the analysis that follows are the adopted
nucleosynthesis sources.  For Type Ia SNe, the yield appears to be independent 
of progenitor model, and hence we adopt the updated Model W7 yields of
Thielemann, Nomoto \& Hashimoto (1993) -- the lower portion of
Table 1 lists the Type Ia yields $<y_{i,\rm SNIa}>$.

\begin{table*}
\caption[]{Average stellar yield in solar masses
for the SNe grids under consideration here
-- T95=Tsujimoto \etal 1995, M92=Maeder 1992, W95=Woosley \&
Weaver 1995, A96=Arnett 1996, LH95=Langer \& Henkel 1995,
TNH=Thielemann et al. 1993. 
Top block - Type II SNe: averaged over the progenitor mass range 
$10\rightarrow 50$ M$_\odot$, for a Salpeter (1955) IMF.  Linear extrapolation
in mass, from the two lowest mass models in the respective grids to
$m=10$ M$_\odot$, assumed.  Bottom block - Type Ia SNe:  mass independent.
}
\begin{center}
\begin{tabular}{lcccccc}
\vspace{2.0mm}
Yield Source  & $<y_{\rm Fe,SNII}>$ & $<y_{\rm O,SNII}>$ & $<y_{\rm
Si,SNII}>$& $<y_{\rm Mg,SNII}>$ & $<y_{\rm Ne,SNII}>$ & $<y_{\rm S,SNII}>$ \\
A96		           & 0.071 & 0.593 &  n/a  & 0.054 & 0.101 &  n/a  \\
T95		           & 0.121 & 1.777 & 0.133 & 0.118 & 0.232 & 0.040 \\
T95+M92		           & 0.121 & 0.923 &  n/a  &  n/a  &  n/a  &  n/a  \\
W95;A;$10^{-4}$Z$_\odot$   & 0.073 & 0.806 & 0.104 & 0.036 & 0.095 & 0.059 \\
W95;B;$10^{-4}$Z$_\odot$   & 0.085 & 1.455 & 0.118 & 0.066 & 0.223 & 0.065 \\
W95;A;Z$_\odot$	           & 0.113 & 1.217 & 0.124 & 0.065 & 0.181 & 0.058 \\
\vspace{2.0mm}
W95;B;Z$_\odot$	           & 0.141 & 1.664 & 0.143 & 0.094 & 0.265 & 0.064 \\
\vspace{2.0mm}
Yield Source  & $<y_{\rm Fe,SNIa}>$ & $<y_{\rm O,SNIa}>$ & $<y_{\rm
Si,SNIa}>$& $<y_{\rm Mg,SNIa}>$ & $<y_{\rm Ne,SNIa}>$ & $<y_{\rm S,SNIa}>$ \\
TNH93  		           & 0.744 & 0.148 & 0.158 & 0.009 & 0.005 & 0.086 \\
\end{tabular}
\end{center}
\end{table*}

\begin{table*}
\caption[]{Comparison of input physics for the five Type II SNe models
discussed herein.  Abbreviations as noted in the caption to Table 1.
Yes/No entries refer
to the inclusion (or not) of pre-SN mass-loss (column 2), unprocessed metals in
the ejecta (column 6), and the effects of explosive nucleosynthesis (column
7).  The $^{12}$C$(\alpha,\gamma)^{16}$O reaction rate is quoted relative to
that of Caughlan \etal (1985) (column 4).  The SN progenitor metallicity
(column 5), intial evolutionary state (\ie pure helium star (He) or zero
age main sequence (ZAMS)) (column 8), and adopted convection treatment
(Sch=Schwarzschild, Led=Ledoux, over=overshooting, semi=semiconvection,
ad=adiabaticity, chem hom=chemical homogeneity, and semi hom=semiconvection
region homogeneity) (column 3) are noted.
}
\begin{center}
\begin{tabular}{lclccccc}
\vspace{2.0mm}
Yield Source  & $\dot{\rm M}$(?) & Convection & $^{12}$C$(\alpha,\gamma)^{16}$O & Z & m$^{\rm unp}$(?) & exp nuc(?) & init state \\
T95	& N & Sch  		& C85  	     & Z$_\odot$    & N & Y & He   \\
M92	& Y & Sch+over		& C85	     & Z$_\odot$/20 & N & N & ZAMS \\
W95    & N & Led+semi          & 0.74$\times$C85 & Z$_\odot$/10 & Y & Y & ZAMS \\
A96     & N & ad+chem hom+semi hom & 0.74$\times$C85 & Z$_\odot$ & Y & N & ZAMS \\
LH95    & Y & $\sim$Led+semi    & 0.74$\times$C85 & Z$_\odot$/10 & Y & N & ZAMS \\
\end{tabular}
\end{center}
\end{table*}

For Type II SNe, the situation is not so clear; there are
large differences in the
predicted yields as a function of mass sequence mass and metallicity -- 
the treatment of convection and mass-loss, in particular, are
highly uncertain.  For what follows, we have chosen a representative sample of
the Type II SNe yields on offer in the 
literature -- (i) Tsujimoto \etal (1995) [`T95' in Tables 1 and 2],
which formed the basis of Ishimaru \& Arimoto's (1997) analysis.  These models
adopt the Schwarzschild criterion for convection, and are not evolved
self-consistently from the zero age main sequence (\ie the models are pure
helium stars, with some assumed He core mass--main sequence mass relation).  The
$^{12}$C$(\alpha,\gamma)^{16}$O reaction rate is that of
Caughlan \etal (1985).  Stellar winds have not been considered in these
models.
(ii) Woosley \& Weaver's (1995) Z=Z$_\odot$ and Z=$10^{-4}$Z$_\odot$ yields, for
both Case A and B SN piston energetics
[`W95;A;Z$_\odot$' and `W95;B;Z$_\odot$', `W95;A;$10^{-4}$Z$_\odot$' and
`W95;B;$10^{-4}$Z$_\odot$']
have been included in our analysis. In their Case B, the final kinetic
ejecta energy was boosted from $\sim 1.2\times 10^{51}$ erg to
$\sim 1.9\times 10^{51}$ erg in order to reduce the effects of
reimplosion of explosively synthesized material.   
Unlike the Tsujimoto \etal (1995) models, this grid has been evolved
self-consistently from the ZAMS, using a $^{12}$C$(\alpha,\gamma)^{16}$O
reaction rate that is $\sim 74$\% that of the
Caughlan \etal (1985) rate used by
Tsujimoto \etal and Thielemann \etal
(1996).
Woosley \& Weaver use the Ledoux convection criterion, with modifications for
semiconvection.  Again, no pre-SN mass-loss is considered in their models.
Woosley \& Weaver (1996) only provide the \it total \rm
mass of a particular element ejected from a star of mass $m$ and metallicity
$Z$ -- including both the newly synthesised material as well as the
initially present, unprocessed, ejecta.  
(iii) Maeder's (1992) oxygen yields are considered in tandem with the iron
yields of Tsujimoto \etal (1995)
[`T95+M92'].  Maeder, along with Langer \& Henkel (1995),
represent the only published grid of yields incorporating mass-loss.  Maeder
uses the Schwarzschild criterion for convection, like Tsujimoto \etal, but also
includes some degree of overshooting.  Oxygen production is severely hampered
in $m\simgt 30$ M$_\odot$ models with solar metallicity,\footnote{The $m\simgt
30$ M$_\odot$, Z=Z$_\odot$, models in the Maeder 1992 grid lose their
hydrogen-rich envelopes and evolve into Wolf-Rayet stars.  Substantial
mass-loss continues through core helium burning, greatly enhancing the helium
and (during later phases) the carbon yields, at the expense of reduced ``fuel''
availability for processing beyond oxygen.} a result that has
profound implications for the Type Ia versus Type II ICM iron argument of
Ishimaru \& Arimoto (1997).  A direct mapping of Maeder's oxygen yields onto 
the Tsujimoto \etal yields, is not optimal by any means, but it does provide
a tantalising clue to the importance of mass-loss consideration.
(iv) Arnett's (1996) oxygen, magnesium, and neon yields have been coupled to
his earlier iron predictions (Arnett 1991) [`A96'].  This has been necessary as the
1996 models contain yields arising from hydrostatic evolution only.  Explosive
nucleosynthesis can modify the silicon, sulphur, and iron results
substantially (Baz\'an 1997, priv comm).  Arnett's new grid of SN models avoids
the use of mixing-length theory with semiconvection (\eg Woosley \& Weaver
1996), and assumes that convection is so efficient that complete adiabaticity
and chemical homogeneity are maintained.  Semiconvective regions are also
assumed to be homogeneous.

By considering Type II SN yields from progenitors with uniform
metallicities, we
have implicitly ignored the effects of galaxy evolution.  
In principle, one ought to couple models of galaxy formation and
chemical evolution directly to the SNII explosion calculations
to determine self-consistent
integrated nucleosynthetic yields. These would be 
superpositions of enrichment from Type II SNe with progenitors
of varying metallicity. By including both high and low WW95
metallcities we do bracket the expected range of average yields 
for that subset of our models, and
provide some indication of the general effects of varying the progenitor 
abundances.

A full discussion of the input physics differences, and their respective
implications, in each of the above SNe models, is well beyond the scope of this
paper.  The reader is strongly urged to turn to
Woosley \& Weaver (1995), Arnett (1996), and Langer\footnote{Figures 1 and 2 in
Langer's (1997) oxygen yield comparison are particularly relevant.  His figure
1 demonstrates the factor of two difference that exists between models with the
\it same \rm convection treatment, while his figure 2 highlights a further
factor of three uncertainty, due to the assumptions regarding semiconvection.
On an even more sombre note, Baz\'an \& Arnett's (1997) multi-dimensional
hydrodynamical simulations of oxygen shell burning during the pre-core collapse
call into question the entire validity of the one-dimensional diffusion-like
algorithms that have been previously used (and are inherently assumed in each
of the yield compilations adopted in this paper)!}
(1997), for details.  For
a specific analysis of the implications for oxygen, the latter reference is
particularly recommended.  Table 2 is our attempt at listing, in as concise a
manner as possible, the 
relevant differences between the Type II SNe yield sources adopted in our
study.

We are now in a position to
quantify the effect of different input physics on our analysis, by
simply examining the predicted $<y_{i,\rm SNII}>$ (recall equation 
\ref{eq:SNIIfe} for each of the yield sources
under consideration);  the upper portion of
Table 1 provides this information for the Salpeter (1955)
IMF noted previously.  While it should be apparent that for some elements 
(\eg Si and S), agreement at the $\sim 50$\% level exists, 
differences of factors of $\sim 2\rightarrow 3$ persist
in the important Fe and O
yields.  These large differences should immediately hint at a potential problem
in quantifying the exact ratio of Type Ia to Type II SNe contribution to the
ICM heavy elements, a point to which we return shortly.

Having determined the mean stellar yields for both Type Ia and II SNe, for the
IMF under consideration, we can now
use equations \ref{eq:ICMfe} and
\ref{eq:SNIIfe} 
(and their analogs for other elements)
to compute the ICM abundance ratios [O,Si,Mg,Ne,S/Fe] as a
function of M$_{\rm Fe,SNIa}/$M$_{\rm Fe,total}$.  The results of this can be
seen in Figure 1 -- the seven different yield source possibilities are labeled,
and the horizontal dotted curve in each frame represents the mean of 
Mushotzky \etal's (1996) ASCA SIS data, as scaled by
Ishimaru \& Arimoto (1997).  Our models adopt the
meteoritic abundance scale of Anders \& Grevesse (1989).

\begin{figure*}
\centering
\vspace{17.0cm}
\includegraphics{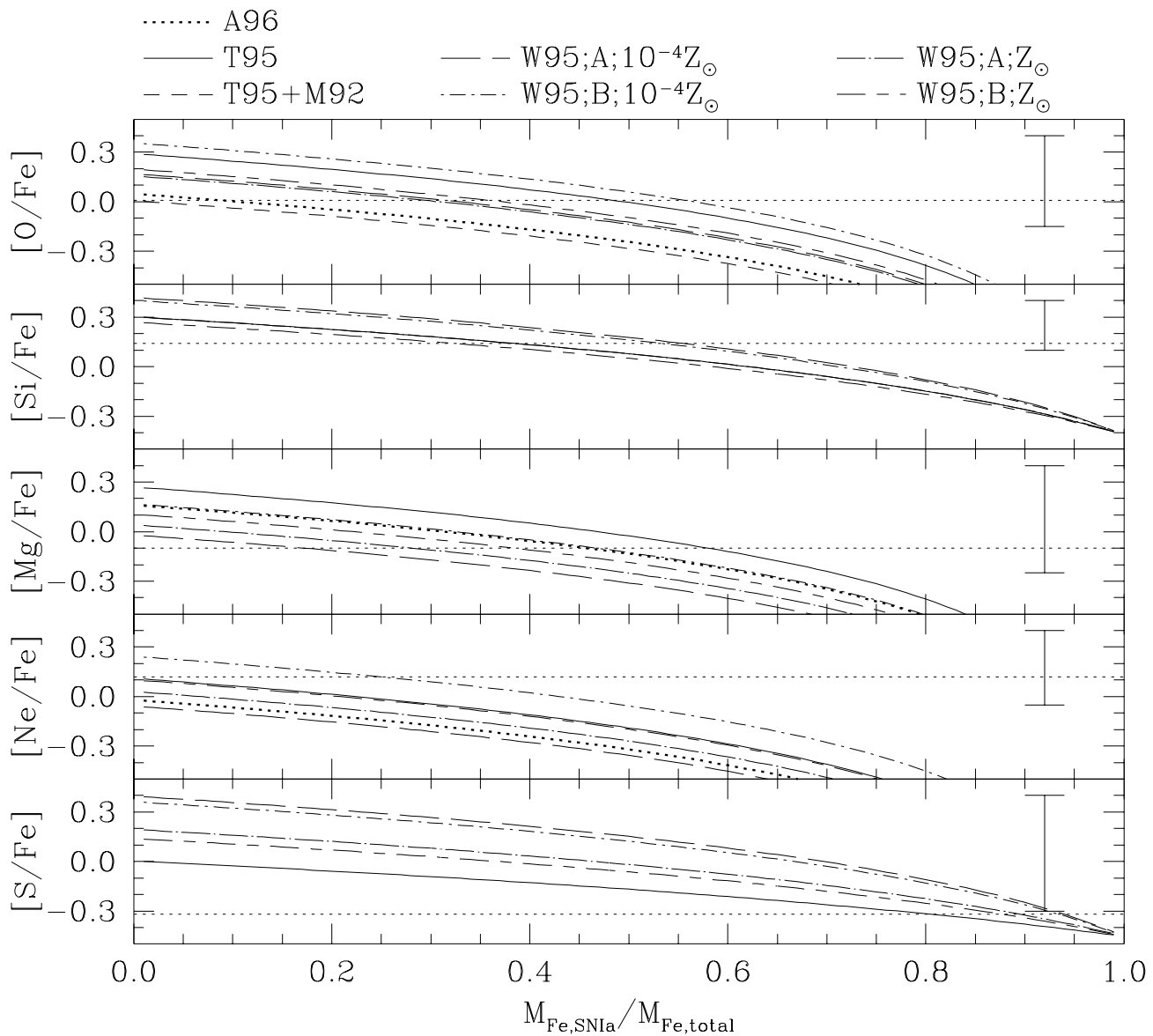}
%\special{psfile=fig1.ps voffset=-120 vscale=88 hscale=88 hoffset=-10}
\caption[]
{Ratio of relative abundance [X$_i$/Fe] as a function of the ICM Type Ia
SNe-originating fraction of Fe.  The Salpeter (1955) IMF ($x=1.35$) was assumed, with
lower and upper mass limits to Type II SNe production of 10 and 50 M$_\odot$,
respectively.  The horizontal dotted curve represents the mean of Mushotzky
{\it et al.}'s (1996) ASCA SIS abundance data, as scaled to the meteoritic
iron abundance by Ishimaru \& Arimoto (1997).  The remaining seven curves
represent the theoretical predictions for seven different Type II SNe yield
scenarios, as described in the text.  The typical error associated with any
individual cluster's ICM elemental abundance determination is indicated in each panel.
\label{fig:fig1}}
\end{figure*}

The solid curves are a reasonable representation of the Ishimaru \& Arimoto
(1997) models - an exact match is difficult as their exact handling of the
yield extrapolation in the $10\rightarrow 13$ M$_\odot$ range
is not provided.  On
the surface, this may seem a trivial matter, but one should
bear in mind that approximately one fifth of the Type II progenitor mass is
tied up in the $10\rightarrow 13$ M$_\odot$ region of the IMF under
consideration.  ``Blind'' linear extrapolation (as adopted here), 
extrapolation from the
$m=13$ M$_\odot$ yields down to effectively zero, or simply setting all yields
for masses below 13 M$_\odot$ to some arbitrary value, all lead to further
factors of $\sim 2$ uncertainties in the results, a point alluded to by
Ishimaru \& Arimoto.

The important result to take from Figure 1 is the intersection of the various
curves with the mean of the ASCA SIS data.  Table 3 provides the
``intersection'' information for each model --
\ie for each element, the fraction of that element originating
from Type Ia SNe (the numerators) required to match 
the mean of the ASCA SIS ICM abundance, and the corresponding
fraction of ICM iron from Type Ia SNe (the denominators).
Because oxygen, silicon, and iron have the best-determined abundances
(Mushotzky \etal 1996), Ishimaru \& Arimoto (1997) base the
bulk of their conclusions upon the [O/Fe] and [Si/Fe] plots, so let us
concentrate on these elements for the time being.

\begin{table*}
\caption[]{
ICM Element Fraction of Type Ia Origin - 
For each entry of the form $a/b$, $a$ represents the mass fraction
of ICM O, Si, Mg, Ne, or S, that originates in Type Ia SNe, for the Type II
yield source under consideration; $b$ represents the 
corresponding Type Ia-originating
ICM Fe fraction.  The $a/b$ pairs shown recover the mean ICM ratios illustrated
in Figure 1.
}
\begin{center}
\begin{tabular}{lccccc}
\vspace{2.0mm}
Yield Source  & O/Fe & Si/Fe & Mg/Fe & Ne/Fe & S/Fe \\
A96		& 0.00/0.08 &   n/a     & 0.01/0.46 & $\,\,$0.00/0.00$^\dagger$ &   n/a    \\
T95		& 0.01/0.48 & 0.11/0.38 & 0.02/0.58 & 0.00/0.00 & 0.58/0.80\\
T95+M92		& 0.00/0.00 &   n/a     &   n/a     &   n/a     &   n/a    \\
W95;A;$10^{-4}$Z$_\odot$	& 0.01/0.31 & 0.15/0.55 & 0.01/0.17 & $\,\,$0.00/0.00$^\dagger$ & 0.69/0.94\\
W95;B;$10^{-4}$Z$_\odot$	& 0.01/0.55 & 0.15/0.53 & 0.01/0.47 & 0.00/0.24 & 0.67/0.93\\
W95;A;Z$_\odot$	& 0.01/0.29 & 0.11/0.38 & 0.01/0.28 & $\,\,$0.00/0.00$^\dagger$ & 0.65/0.89\\
W95;B;Z$_\odot$	& 0.01/0.35 & 0.09/0.32 & 0.01/0.38 & 0.00/0.00 & 0.63/0.87\\
\end{tabular}
\end{center}
\noindent
{\footnotesize
$^\dagger$\it No \rm mixture of Type Ia and II SNe can recover the observed ICM
ratios, for the yield source in question.}
\end{table*}

From the solid curves in the top two panels of
Figure 1 (or the corresponding entries in Table 3), 
we can see that Ishimaru \& Arimoto (1997)
favour a Type Ia iron fractional contribution to the
ICM of $\sim 48$\% and $\sim 38$\%, based upon oxygen and silicon,
respectively, because of their use of the Tsujimoto \etal (1995) yields
\footnote{We should note
though that if the massive star
IMF slope is closer to the Scalo (1986) value of $x\approx 1.7$, as opposed to
the Salpeter (1955) value of $x\approx 1.35$, then the implied 
ICM M$_{\rm Fe,SNIa}/$M$_{\rm Fe,SNII}$ ratios for oxygen and silicon would be
reduced from $\sim 0.48$ and $\sim 0.38$, respectively, to $\sim 0.32$ and $\sim
0.26$, respectively, thereby strengthening the support for a $\sim 3\rightarrow
4\times$ Type II-to-Type Ia predominance ratio.}.
Simply replacing the Tsujimoto \etal yields with those of
Woosley \& Weaver (1995) would allow
anything in the range of $\sim 30\rightarrow 55$\%, depending upon assumed SNe
energetics (an important consideration in determining the amount of Si, S,
and Fe, for example, that fall back onto the collapsed
remnant) and progenitor
metallicity.

Of particular interest is the behaviour of the [O/Fe] models when the effects 
of Maeder (1992)-style mass-loss or Arnett (1996)-style convection 
are taken into account.  
Here we are somewhat restricted in that both these grids only
consider hydrostatic evolution; the effects of explosive
nucleosynthesis upon the yields are not included, and these
can be profound for Si, S, and Fe.  Thusly, we simply examine O (for
the Maeder grid) and O, Mg, and Ne (for the Arnett grid), adopting reasonable
assumptions for the Fe for each grid -- Tsujimoto \etal 1995, for Maeder
(because of some convection-treatment similarities), and
Arnett 1991, for Arnett (for obvious reasons).

Perhaps the most important result to take away from our study is the behaviour
of these Arnett (1996) and Maeder (1992) models 
(A96 and T95+M92, respectively)
in Figure 1 (and Table 3).
For these Type II SNe yield sources, any model
that includes an ICM Type Ia iron fractional contribution $\simgt 5$\% is at
odds with the mean of the ASCA SIS cluster data.\rm  This appears to be a
fairly robust conclusion -- the most modern treatments of convection and
mass-loss \it both \rm act in the direction of favouring a dominant Type II
origin to the ICM iron abundance.
\footnote{We have not shown the behaviour of Langer \& Henkel's (1995) oxygen
yields, calculated in the presence of pre-SN mass-loss, 
coupled to the iron yields of Woosley \& Weaver (1995), to whom their
convection treatment most closely resembles.  
The former's predictions regarding the
mass of ejected oxygen parallels almost exactly that of Arnett (1996),
and their mapping onto the
models of Woosley \& Weaver (1995) would likewise tend to
indicate a dominant role for Type II SNe in ICM enrichment --
analogous to the case of the M92 extension of the Tsujimoto et al.
(1995) models.}

Unfortunately, while oxygen may be pointing toward a Type II SNe origin to
the ICM abundances, the situation with
silicon (the other element that Ishimaru \& Arimoto 1997
considered as the most important in their analysis) is not so obvious.  
An ICM Type Ia iron fractional contribution in the
range $\sim 30\rightarrow 55$\%
is allowed by the current grid of Type II SNe models, but we cannot, as of yet,
quantify accurately
the effects of mass-loss and other convection treatments (a la Maeder
1992 and Arnett 1996) on the silicon yields, as both these models only considered
hydrostatic evolution.  Explosive nucleosynthesis computations for these grids are
eagerly anticipated.

While Mg, Ne, and S are usually given less weight in these ICM abundance
analyses (see Mushotzky \etal 1996 for a discussion of the inherent
difficulties in accurately determining their respective abundances), their
behaviour in the $[{\rm X}_i/{\rm Fe}]-{\rm M}_{\rm Fe,SNIa}/{\rm M}_{\rm
Fe,total}$ plane (Figure 1) is nonetheless interesting.  (i) The Tsujimoto
\etal (1995) [Mg/Fe] ratios are significantly higher than those of Woosley \&
Weaver (1995) and, to a lesser extent, Arnett (1996).  While adopting the
latter would lead one to conclude that Type Ia SNe contribute $\sim
15\rightarrow 45$\% of the ICM Fe, Ishimaru \& Arimoto (1997), because they
followed Tsujimoto \etal, favour $\sim 60$\%. (ii) There is an apparent dichotomy
between the neon and sulphur results.  One must be 
cautious against
putting too much weight on the neon observations, due to potential
systematic problems (Mushotzky \etal 1996).  On the other hand, as already
commented upon in Mushotzky \etal (1996) and Loewenstein \& Mushotzky (1996),
there would appear to be no escape from the fact that the sulphur abundances
are at odds with a Type II-dominated origin to the ICM iron abundance
(although the uncertainties are large).  This
appears to be entirely independent of adopted Type II SNe yields.

Finally, while many arguments still remain unresolved as far as the ICM Type Ia
SNe iron fraction goes, it should be readily apparent from inspection of Table 3
that regardless of Type II yield compilation assumed, $\sim 90\rightarrow
100$\% of the ICM oxygen, silicon, magnesium, and neon originated from Type II
SNe. Moreover, since Type Ia and Type II SN kinetic energies are similar
but the average
Fe yield 5-10 times higher for Type Ia, Type II SNe must dominate
the energetics of early galactic winds -- even if half of the Fe originates
in Type I SNe.

\section{Summary}
\label{summary}

Ishimaru \& Arimoto (1997) recently put the four
Mushotzky \etal (1996) ASCA ICM determinations onto the meteoritic abundance
scale and concluded that $\simgt 50$\% of the ICM iron must have originated
from Type Ia SNe.  This fractional value is linked inexorably to the adopted
Type II SNe yields used in their analysis -- \ie those of Tsujimoto \etal
(1995).

We have re-examined the Ishimaru \& Arimoto results, adopting a range of the
Type II SNe models available in the literature.  These models sample a wide
variety of input physics (\eg pre-SN stellar winds, convective overshooting,
reaction rates) and provide, for the first time, a better appreciation for the
dependence of Type Ia versus Type II arguments on the adopted yields.  
Specifically, we have demonstrated that the most recent treatments of convection
(Arnet 1996) and mass-loss (Maeder 1992 and Langer \& Henkel 1995) 
\it both reduce the ICM
Type Ia iron fractional contribution from $\simgt 50$\% to $\simlt 5$\%. \rm
This is a preliminary result that needs confirmation once full grids of Type 
II SNe yields are published that include self-consistent treatments of 
mass-loss, explosive nucleosynthesis, fall-back onto the remnant,
and large atomic networks.

Several further caveats should be expressed -- (i) there is no \it a
priori \rm reason why the Type II progenitors need follow an $x=1.35$ IMF
power-law slope; as noted earlier, if the slope more closely resembles a Scalo
(1986) IMF ($x\approx 1.7$), then the ICM Type Ia iron fraction is reduced a
further $\sim 40$\%.  (ii)  We have not been concerned with cluster-to-cluster
differences in the effective Type Ia versus Type II contributions, only the
unweighted mean.  (iii)  There are still fairly large statistical uncertainties
in the abundance determinations (\eg uncertainties of $\sim 25$\%, $\sim 50$\%,
$\sim 50$\%, $\sim 100$\%, and $\sim 100$\%, for Si/Fe, O/Fe, Ne/Fe, Mg/Fe, and
S/Fe, respectively).  (iv) All Type Ia yields are tied exclusively to
the W7 model of Thielemann \etal (1993). (v) The $10\rightarrow 13$ M$_\odot$
yields are essentially unknown; we have extrapolated from the lower-mass limits
in the various grids, but there is no reason to suspect that this is 
entirely satisfactory.  For the IMFs discussed in this paper, $\sim
20\rightarrow 25$\% of the mass in a stellar generation is locked up in this
uncertain regime.

In conclusion, while our model predictions for the ICM [O/Fe] appear to favour
a highly dominant Type II SNe origin to the ICM iron, especially when coupled
with the Maeder (1992), or Arnett (1996) oxygen
yields, we cannot unequivocally state that this is the case -- there still
exists much uncertainty in the massive star models (in particular, convection
mass-loss, reaction rates, location of the iron ``cut'', and
fall-back onto the remnant).  Regardless of the lack of a definitive answer as
to the Type Ia versus Type II ICM iron origin, this study does illustrate its
underappreciated sensitivity to the adopted Type II SNe yields.

\section*{ACKNOWLEDGEMENTS}
BKG acknowledges the financial assistance of NSERC, through its Postdoctoral
Fellowship program.  We thank Grant Baz\'an 
and Una Hwang for helpful discussions.

\end{document}